\documentclass[prb,aps,12pt,a4paper,showpacs]{revtex4}
\usepackage{epsfig}
\begin{document}
\title{Effective Dielectric Constants of Photonic Crystal of Aligned Anisotropic Cylinders: Application to the Optical Response of
 Periodic Array of Carbon Nanotubes}
\author{ E.~Reyes$^1$, A.A.~Krokhin$^{2}$, and J. Roberts$^{2}$ }
\address{$^1$Instituto de F{\'\i}sica, Universidad Aut\'onoma de Puebla, Apartado Postal J-48,
Puebla, 72570 Mexico}
\address{$^2$Department of Physics, University of North Texas, P.O. Box 311427, Denton, TX 76203}
\date{\today}
\begin{abstract}
We calculate the static dielectric tensor of a periodic system of
aligned anisotropic dielectric cylinders. Exact analytical
formulas for the effective dielectric constants for the $E$- and
$H$- eigenmodes are obtained for arbitrary 2D Bravais lattice and
arbitrary cross-section of anisotropic cylinders. It is shown that
depending on the symmetry of the unit cell photonic crystal of
anisotropic cylinders behaves in the low-frequency limit like
uniaxial or biaxial natural crystal. The developed theory of
homogenization of anisotropic cylinders is applied for
calculations of the dielectric properties of photonic crystals of
carbon nanotubes.
\end{abstract}
\date{\today}
\maketitle
{42.70.Qs, 41.20.Jb, 42.25.Lc}

\section{Introduction}
Periodic dielectric structures - photonic crystals (PC) - have
found many useful technological applications.\cite{Noda} Progress
in the development of optical devices that operate using the
principles of the control of light as proposed by
Yablonovich\cite{Yablo} gave rise to theoretical studies of the
general properties of the spectra of PC's.\cite{Kuch} In
particular, the region of low frequencies, where the ideas and
methods of the theory of homogenization\cite{Hom} are applicable,
has attracted a significant amount of attention in the last
decade. In the low-frequency limit the light wave in a
non-absorbable periodic medium has linear dispersion, $\omega
\propto k$. This allows the replacement a real inhomogeneous medium by an
effective homogeneous one with dielectric permittivity
\begin{equation}
\label{one} \epsilon_{eff}(\hat{\bf k}) =\lim _{k \rightarrow 0
}\, \left(\frac{ck}{\omega} \right)^{2}.
\end{equation}
In the general case, this effective parameter depends on the
direction of propagation $\hat{\bf k}={\bf k}/k$ and has tensor
structure. The latter property is emphasized for 2D PC's, which
are anisotropic uniaxial or biaxial crystals, depending on the
symmetry of the unit cell.\cite{Halev} Unlike this, 3D PC's may be
isotropic.\cite{Datta}

Optical anisotropy of the PC's studied in Refs.
[\cite{Halev,Datta,Arka01}] is determined by the geometry of the
unit cell only. The constituents themselves are considered to be
isotropic dielectrics. This is not the case, for example, for a
structure of aligned carbon nanotubes arranged periodically in the
plane perpendicular to the tubes. Here anisotropy manifests itself
at the "microscopic" level, since the nanotubes ("atoms" of the
PC) possess a natural anisotropy. This anisotropy originates from
the layered structure of the crystal of graphite, which has
different dielectric constants along the $c$-axis and in the
perpendicular plane. The static values of these dielectric
constants are $\varepsilon_{\parallel} = 1.8225$ and
$\varepsilon_{\perp}  = 5.226$.\cite{Hand} The elongated topology
and the natural anisotropy of graphite cause PC's of carbon
nanotubes to exhibit large optical anisotropy.\cite{Heer}
Three-dimensional PC's with anisotropic dielectric spheres have
been studied in Refs.[\cite{Zabel}]. It was shown, that, depending
on the symmetry of the unit cell, the anisotropy of the spheres
may be favorable for either broadening or narrowing the band gaps.

High anisotropy of 2D photonic crystals may find interesting
applications in nanophotonics as it was recently proposed by
Artigas and Torner.\cite{Tor} Namely, the surface of an anisotropic
2D photonic crystal supports propagation of a surface wave predicted
by Dyakonov.\cite{Dyak} The surface mode does not radiate and is
localized close to the surface due to the interference between the
ordinary and extraordinary waves. In natural crystals, it can be
hardly observed because of the low anisotropy. Since it is a surface wave
with very low energy losses, the Dyakonov wave may replace surface
plasmons in the near-field optics and integrated photonic
circuits.

An effective medium theory for the PC of anisotropic carbon
nanotubes was proposed in Ref. [\cite{Garcia}]. This theory is
based on the Maxwell-Garnet approximation.\cite{Max} and it leads
to a simple analytical formula for the effective dielectric
constant. The formula is valid at low filling fractions. Since the
Maxwell-Garnett approximation does not discriminate between
periodic and non-periodic arrangements of the nanotubes, the
in-plane anisotropy of the effective dielectric constant is lost.
In other words, $\varepsilon_{eff}$ in Eq. (\ref{one}) is
independent on $\hat{\bf k}$, i.e., the Maxwell-Garnet
approximation always leads to an effective medium which is
equivalent to a uniaxial crystal. The PC considered in Ref.
[\cite{Garcia}] has a square unit cell that gives rise to in-plane
isotropy of the effective dielectric tensor.

In this paper we extend the results of the theory of
homogenization\cite{Halev} to the case of anisotropic dielectric
cylinders. Exact analytical formulas are obtained for the
principal effective dielectric constants of a 2D PC with an
arbitrary cross-sectional form of anisotropic cylinders, arbitrary
Bravais lattice, and filling fraction. We compare our results with
the results obtained for the PC of carbon nanotubes in the
modified Maxwell-Garnet approximation.\cite{Garcia} This
comparison shows that the Maxwell-Garnett approximation gives
correct results for a dilute system with filling fraction less
then $5\%$. However, at higher filling fractions the
Maxwell-Garnet approximation overestimates the values for the
effective dielectric constant. For a close-packed structure the
error is about $10\%$. We also consider a PC with a  rectangular
unit cell and calculate two different in-plane dielectric
constants. In this case the corresponding effective medium is a
biaxial crystal. It was argued that for the $H$-polarized mode the
effective dielectric constant for hollow and solid cylinders are
practically indistinguishable.\cite{Garcia} Using our approach, we
study the effect of the internal cavity on the effective
dielectric constant and show explicitly how the effective
dielectric constant decreases with the internal radius.

\section{The Fourier expansion method in the long-wavelength limit}
We consider a 2D periodic structure of dielectric cylinders with
their axes parallel to $z$ and whose cross section can have an
arbitrary shape. The cylinders are imbedded in a dielectric
matrix. A 2D PC supports propagation of two uncoupled modes with
either $E$-polarization (where the vector ${\bf E}$ is parallel to
the cylinders), or $H$-polarization (in this case the vector ${\bf
H}$ is parallel to the cylinders). The background material is an
isotropic dielectric with permittivity $\varepsilon_b$ and the
cylinders are rolled up from an anisotropic dielectric sheet
characterized by a tensor $\hat\varepsilon^{(a)}$. For carbon
nanotubes, this tensor has two different eigenvalues, and in
cylindrical coordinates is represented by a diagonal matrix with
elements $\varepsilon^{(a)}_{\theta\theta}=\varepsilon^{(a)}_{zz}=
\varepsilon_{\perp}$ and
$\varepsilon^{(a)}_{rr}=\varepsilon_{\parallel}$. As a whole, the
periodic inhomogeneous dielectric medium is characterized by the
coordinate-dependent dielectric tensor,
\begin{equation}
\label{two}
\mathbf{\hat \varepsilon}({\bf r})=
\left( \begin{array}{ccc}
\varepsilon_{xx}({\bf r}) & \varepsilon_{xy}({\bf r}) & 0 \\
\varepsilon_{yx}({\bf r}) & \varepsilon_{yy}({\bf r}) & 0 \\
0 & 0 & \varepsilon_{zz}({\bf r})
\end{array} \right).
\end{equation}
Inside the cylinders this tensor coincides with
$\hat\varepsilon^{(a)}$ and outside the cylinders it reduces to a
scalar, $\varepsilon_{b}\delta_{ik}$.

The wave equations for the $E$- and $H$-polarized modes with
frequency $\omega$ have the following form:
\begin{equation}
\label{three} \nabla^{2} E  =
\frac{\omega^2}{c^2}\varepsilon_{zz}({\bf r}) E,
\end{equation}
\begin{equation}
\label{four} \frac{\partial}{\partial
x_{i}}\left(a_{ji}\frac{\partial H}{\partial x_j} \right) +
\frac{\omega^2}{c^2} H =0, \,\,\,\,\,\,\, x_i,x_j = x,y.
\end{equation}
Here $E=E(x,y)$ and $H=H(x,y)$ are the amplitudes of the $E$ and
$H$ monochromatic eigenmodes, respectively, and $a_{ij}$ is a
$2\times2$ Hermitian matrix with determinant 1:
\begin{equation}
\label{ten}  a_{ij}({\bf r})=
 \frac{1}{\varepsilon_{xx}({\bf r})\varepsilon_{yy}({\bf r})-\varepsilon_{xy}({\bf r})\varepsilon_{yx}({\bf r})}\left(
\begin{array}{cc}
\varepsilon_{xx}({\bf r}) & \varepsilon_{xy}({\bf r}) \\
\varepsilon_{yx}({\bf r}) & \varepsilon_{yy}({\bf r}) \\
\end{array} \right).
\end{equation}
The determinant in the right-hand side can be written as a product
of two eigenvalues. Within the graphite wall of the cylinders this
product is $\varepsilon_{\parallel}\varepsilon_{\perp}$. Outside
the wall it is either $(\varepsilon^{(b)})^2$ (for  $\bf r$ being
outside the cylinders) or 1 for  the interior region of the
cylinder, which is free from the dielectric material.

Eq.(\ref{three}) depends only on the $zz$ component of the
dielectric tensor, i.e., it is insensitive to the in-plane
anisotropy. Therefore the effective dielectric constant in the
long-wavelength limit is the same as for the parallel arrangement
of isotropic cylinders (not necessarily periodic). It is given by
a simple formula
\begin{equation}
\label{five}
\varepsilon^{(E)}_{eff} = \bar\varepsilon_{zz},
\end{equation}
where
\begin{equation}
\label{six} \bar\varepsilon_{zz} =
\frac{1}{A_c}\int_{{A_c}}\varepsilon_{zz}({\bf r})d{\bf r},
\end{equation}
is the average over the area $A_c$ of the unit cell $zz$ component
of the tensor (\ref{two}). For a binary composite
$\varepsilon^{(E)}_{eff} = \bar\varepsilon_{zz} = f
\varepsilon_{\perp} + (1-f)\varepsilon_{b}$, where $f$ is the
filling fraction of the component $a$.

To obtain the long-wavelength limit for Eq. (\ref{four}) we apply
the method of plane waves.\cite{Halev} Using the Bloch theorem and
the periodicity of the function $a_{ij}({\bf r})$, we get the
Fourier expansions,
\begin{equation}
\label{seven} H({\bf r})= \exp{(i{\bf k}\cdot {\bf r})}\sum_{\bf
G}{h_{\bf k}({\bf G})\exp{(i{\bf G}\cdot {\bf r}})},
\end{equation}
\begin{equation}
\label{seven'} a_{ij}({\bf r})= \sum_{\bf G}{a_{ij}({\bf
G})\exp{(i{\bf G}\cdot {\bf r}})},
\end{equation}
where the Fourier coefficients $a_{ij}({\bf G})$ are given by
\begin{equation}
\label{eight}
\begin{array}{cc}
 a_{ij}({\bf G})=  \frac{1}{A_c} \int_{{A_c}}a_{ij}({\bf r})
\exp(-i{\bf G \cdot r})d{\bf r} \\
=\frac{1}{\varepsilon_{\parallel}\varepsilon_{\perp}
A_c}\int_{{a}}\varepsilon^{(a)}_{ij}({\bf r}) \exp(-i{\bf G \cdot
r})d{\bf r} +
\frac{\delta_{ij}}{A_c}\left[\frac{1}{\varepsilon_b^{2}}\int_{{b}}
\exp(-i{\bf G \cdot r})d{\bf r} + \int_{{c}} \exp(-i{\bf G \cdot
r})d{\bf r}\right]. \nonumber
\end{array}
\end{equation}
Here $\bf G$ gives the reciprocal-lattice vectors. Indices $a$,
$b$, and $c$ at the integrals label the domains of integration --
within the graphite walls, within the dielectric matrix,  and
within the interior of the hollow cylinders respectively.
Substituting Eqs. (\ref{seven}) and (\ref{seven'}) into Eq.
(\ref{four}) we get a generalized eigenvalue problem in $\bf
G$-space,
\begin{equation}
\label{nine}
 \sum_{\bf G^{\prime}}{ a_{ij} ({\bf G} - {\bf G^{\prime}}) ({\bf k}+{\bf G})_{i}} ( {\bf k} + {\bf G^{\prime}}
)_{j} \,h_{\bf k}({\bf G}^{\prime}) = (\omega^2/c^2) h_{\bf
k}({\bf G}), \,\,\,\, i,j =x,y.
\end{equation}

Eq. (\ref{nine}) is  an infinite set of homogeneous linear
equations for the eigenfunctions $h_{\bf k}({\bf G})$. The
dispersion relation $\omega = \omega_n(\bf k)$ ($n=1,2,\dots$) is
obtained from the condition that this set has non-trivial
solutions.

The periodic medium behaves as a homogeneous one if the Bloch wave
(\ref{seven}) approaches a plane wave. This occurs if the Fourier
coefficients with ${\bf G} \neq 0$ in (\ref{seven}) vanish in the
long-wavelength limit. To obtain the behavior of $h_{\bf k}({\bf
G})$ we substitute ${\bf G} =0$ in both sides of Eq. (\ref{nine}),
divide the both sides by $h_{\bf k}({\bf G}=0)$ and take the limit
as $k \rightarrow 0$
\begin{equation}
\label{eleven}
1 =  \frac{1}{\omega ^2/c^2-\overline{a}_{ij}k_{i}k_{j}}%
\sum_{\bf{G}^{\prime }\neq 0} a_{ij} (-{\bf G}^{\prime }) \, k_i
\,{G^{\prime }}_j \,h^{\star}_{\bf k} {\bf (G^{\prime })} \,.
\end{equation}
Here $\bar a_{ij} \equiv  a_{ij}({\bf G}=0)$ is the bulk average
of the matrix (\ref{ten}) and $h^{\star}_{\bf k} {\bf (G)} =
h_{\bf k} {\bf (G)}/h_{\bf k}({\bf G}=0)$ is the normalized
Fourier coefficient. In the long-wavelength limit the coefficients
of $h_{\bf k}^{\star}({\bf G}^{\prime})$ in the right-hand side
are inversely proportional to $k$. In order to make the sum
finite, the amplitudes of non-zero harmonics, $h_{\bf
k}^{\star}({\bf G}^{\prime})$ must approach zero linearly with
$k$. Thus, in the long-wavelength limit the Bloch wave
(\ref{seven}) can be written as a linear expansion over powers of
$k$:
\begin{equation}
\label{twelve} H({\bf r})= \exp{(i{\bf k}\cdot {\bf r})}\left[
h_0+ \sum_{{\bf G} \neq 0}{ h^{\star}_{\bf k}({\bf G})\exp{(i{\bf
G}\cdot {\bf r}})}\right].
\end{equation}
Since the sum over $\bf G$ vanishes linearly with $k$, Eq.
(\ref{twelve}) shows that the medium becomes homogeneous, i.e. the
solution of the wave equation (\ref{four}) approaches a plane wave
with an amplitude $h_0=h_{\bf k}({\bf G}=0)$ when $k\rightarrow
0$.

Now, to calculate the effective dielectric constant (\ref{one}),
we develop a theory of perturbation with respect to a small
parameter $ka$ ($a$ is the lattice period). In Eq. (\ref{nine}) we
keep the linear terms and obtain the following relation,
\begin{equation}
\label{thirteen} a_{ij}( {\bf G}) G_{i} k_{j} + \sum_{\bf
G^{\prime}\neq 0}{ a_{ij}({\bf G} - {\bf G^{\prime}}) G_{i}
G^{\prime}_{j}\, h_{\bf k}^{\star}({\bf G^{\prime}})} = 0.
\end{equation}
The quadratic approximation is given by Eq. (\ref{eleven}), which
gives another linear relation between the eigenvectors $h_{\bf
k}^{\star}({\bf G})$. Note that this relation is obtained from the
eigenvalue problem Eq. (\ref{nine}) for ${\bf G} = 0$ and the
linear approximation Eq. (\ref{thirteen}) is obtained for ${\bf G}
\neq 0$. The linear relations, Eqs. (\ref{eleven}) and
(\ref{thirteen}), are the homogenized equations for the Fourier
components of the magnetic field. These equations are consistent,
if the corresponding determinant vanishes:
\begin{equation}
\det_{ {\bf{G,G}^{\prime }}\neq 0} \left[ \Lambda a_{ij}({\bf G} -
{\bf G^{\prime}}) G_{i} G^{\prime}_{j}  - a_{ij}({\bf G}) a_{mn}(
- {\bf G^{\prime}})\,  G_{i}\,  n_{j}  n_{m} \,G^{\prime}_{n}
\right] =0. \label{sixteen}
\end{equation}
Here ${\bf n} = {\bf k}/k$ is the unit vector in the direction of
propagation and $\Lambda = (\bar a_{ij} n_{i}  n_{j} -
\varepsilon_{eff}^{-1})$. Since Eqs. (\ref{eleven}) and
(\ref{thirteen}) are homogenous with respect to $k$, the
dispersion equation (\ref{sixteen}) depends only on the inverse
effective dielectric constant, $(\omega/ck)^2$. This fact is a
manifestation of a general property: At low frequencies an
electromagnetic wave has a linear dispersion in a periodic
dielectric medium.

Although Eq. (\ref{sixteen}) is an infinite-order polynomial
equation in $\Lambda $, it turns out that it has only a {\it
unique} nonzero solution. The fact that the second term in the
determinant Eq. (\ref{sixteen}) is a product of two factors, one
of which depends only on $\bf G$ and the other only on $\bf
G^{\prime}$, plays a crucial role. Omitting the mathematical
details, which can be found in Ref. [\cite{Halev}], we obtain the
final answer for the inverse effective dielectric constant
obtained from Eq. (\ref{sixteen}) as:
\begin{equation}
\frac{1}{\varepsilon^{(H)}_{eff}(\hat{\bf n})}=  \bar a_{ij} n_{i}
n_{j} - \sum_{\bf{G,G}^{\prime }\neq 0}{a_{ij}\,({\bf G}) a_{mn}(-
{\bf G^{\prime }}) \, n_{j}\, G_{i} \,  n_{m}\, G^{\prime}_{n}}
\left[ a_{kl}({\bf G} - {\bf G^{\prime}})\, G_{k}\, G^{\prime
}_{l} \right]^{-1}. \label{nineteen}
\end{equation}
Here $[\dots]^{-1}$ implies the inverse matrix in $\bf G$-space.
Eq. (\ref{nineteen}) is valid for an arbitrary form of the unit
cell, geometry of the cylindrical inclusions, material composition
of the photonic crystal, and the direction of propagation in the
plane of periodicity. In the case when $a_{ij}({\bf G}) =
\eta({\bf G})\, \delta_{ij}$ Eq. (\ref{nineteen}) is reduced to
the formula obtained for isotropic cylinders.\cite{Halev}

\section{Index ellipsoid}

As any natural crystal, artificial PC in the long-wavelength limit
can be characterized by an index ellipsoid.\cite{BW} Taking into
account Eq. (\ref{five}) the equation for this ellipsoid can be
written as follows:
\begin{equation}
\label{index} \frac{x_0^2}{\varepsilon_1} +
\frac{y_0^2}{\varepsilon_2} +
\frac{z_0^2}{\bar{\varepsilon}_{zz}}=1.
\end{equation}
Here $x_0,y_0,z_0$ are three mutually orthogonal directions along
which the vectors of the electric field, $\bf E$, and of the
displacement, $\bf D$, are parallel to each other. For the
$E$-mode we have $\bf E
\parallel D
\parallel \hat z$, i.e. the $z_0$ direction coincides with
$z$-axis. In the $x-y$ plane the cross section of the index
ellipsoid is given by Eq. (\ref{nineteen}), which can be rewritten
in the canonical form as
\begin{equation}
\frac 1{\varepsilon ^{(H)}_{eff}(\varphi )}=({\bar a}_{xx}
-A_{xx})\cos ^2\varphi   + ({\bar {a}_{yy} }-A_{yy})\sin ^2\varphi
+({\bar {a}_{xy}}-A_{xy})\sin 2\varphi . \label{twenty}
\end{equation}
Here
 \begin{equation}
A_{ij}=\sum_{{\bf {G,G}^{\prime }}\neq 0} a_{ik}({\bf
G})\,a_{jl}({-\bf G}^{\prime})G_k G_l^{\,\prime }
 \left[ a_{mn} ({\bf {G}^{\prime }-G}) G_m G_n^\prime \right]^{-1}, \hspace{0.5cm}
 i,j,k,l,m,n  =x,y\,.
\label{twentyone}
\end{equation}
Eq. (\ref{twenty}) describes a rotated ellipse in the polar
coordinates $(\rho,\varphi)$. The radius $\rho (\varphi )=
\sqrt{\varepsilon^{(H)}_{eff}(\varphi)}$ gives the index of
refraction of $H$-mode and the angle $\varphi$ is related to the
direction of propagation, ${\bf \hat n} = (\cos\varphi,
\sin\varphi)$. The directions $x_0$ and $y_0$ coincide with the
semi-axes of the ellipse given by Eq. (\ref{twenty}) and the
in-plane indices of refraction $\sqrt\varepsilon_1$,
$\sqrt\varepsilon_2$ are given by the lengths of the semi-axes.
The angle of rotation of the axes of the ellipse Eq.
({\ref{twenty}) with respect to the arbitrary axes $x,y$, is given
by the relation
\begin{equation}
\tan(2 \theta)=\frac{2A_{xy}}{A_{yy}-A_{xx}}.
\label{twentythree}
\end{equation}

If the unit cell possesses a third- or higher-order rotational
axis $z$, then the tensor $A_{ij}$ is reduced to a scalar,
$A_{ij}=A\delta_{ij}$, and the ellipse given by Eq. (\ref{twenty})
is reduced to a circle. In this case the PC in the long-wavelength
limit behaves like a uniaxial crystal; otherwise, it is biaxial.

\section{Uniaxial and biaxial PC's of solid graphite cylinders}

In this section, we study 2D PC of solid carbon cylinders arranged
in square and rectangular lattices. In Cartesian coordinates the
dielectric function of a carbon cylinder is given by the
tensor\cite{Garcia}
\begin{equation}
\label{twentyfour} \mathbf{\hat \varepsilon}^{(a)}({\bf r})=
\left(
\begin{array}{ccc}
\frac{x^2}{r^2}\varepsilon_{\parallel}+\frac{y^2}{r^2}\varepsilon_{\perp}
& \frac{xy}{r^2}(\varepsilon_{\parallel}-
\varepsilon_{\perp})& 0 \\
\frac{xy}{r^2}(\varepsilon_{\parallel}-\varepsilon_{\perp}) &
\frac{y^2}{r^2}\varepsilon_{\parallel}+\frac{x^2}{r^2}
\varepsilon_{\perp} & 0 \\
0 & 0 & \varepsilon_{\perp}
\end{array} \right).
\end{equation}
In numerical calculations we use Eq. (\ref{twenty}) and
(\ref{twentyone}). For rectangular and square lattices with
circular cylinders the semi-axes of the index ellipsoid are
directed along the basic lattice vectors. Because of the high symmetry of
the unit cell, the off-diagonal elements of the tensor $a_{ik}(\bf
G)$ vanish. The diagonal elements for hollow cylinders with outer
and inner radii $ R$ and $\gamma R$, respectively, ($0\leq
\gamma\leq 1$)  have the following form:
\begin{equation}
\label{twentyfive}
 a_{xx}({\bf G})= \left\{
\begin{array}{ll}
\frac{\pi R^2}{A_c}
\left[\frac{1}{2}\left(\varepsilon_{\parallel}^{-1} +
\varepsilon_{\perp}^{-1}\right) (1-\gamma^2)\right] +
\varepsilon_b^{-1}(1-\frac{\pi
R^2}{A_c}(1-\gamma^2))   , & {\bf G} = 0, \\
\frac{2\pi}{A_c
G^2}\left\{GR\left(\varepsilon_{\perp}^{-1}-\varepsilon_b^{-1}\right)\left[J_1({GR})
-\gamma J_1(\gamma GR)\right] \right. \nonumber\\
 +\left. \,\left(\varepsilon_{\parallel}^{-1}-\varepsilon_{\perp}^{-1}\right)\left[J_0({GR})-J_0(\gamma
GR)\right] \right\},  & {\bf G} \neq 0 \, .
\end{array} \right.
\end{equation}
The diagonal element  $a_{yy}(\bf G)$ is obtained from Eq.
(\ref{twentyfive}) by the replacement $\varepsilon_{\perp}
\leftrightarrow \varepsilon_{\parallel} $.

In this section we consider solid cylinders, i.e. $\gamma \equiv
0$. The circles in Fig. \ref{fig.1} show the effective dielectric
constant, given by Eq. (\ref{twenty}), of the $H$-mode as a
function of the filling fraction, $f= \pi R^2/A_c$, for uniaxial
PC with a square lattice. The number of $\bf G$ values (plane
waves) considered in this calculation was 1200, which provided a
good convergence in Eq. (\ref{twenty}). The dielectric constant
for the extra-ordinary mode ($E$-mode), Eq. (\ref{five}) (shown by
triangles in Fig. \ref{fig.1}) is always larger than that for the
ordinary wave ($H$-mode). Therefore, the effective medium is a
uniaxial $\it positive$ optically anisotropic crystal.
\begin{figure}
\begin{center}
\vspace{-0.5cm} \hspace {-0.5cm} \epsfysize 8cm \epsfbox{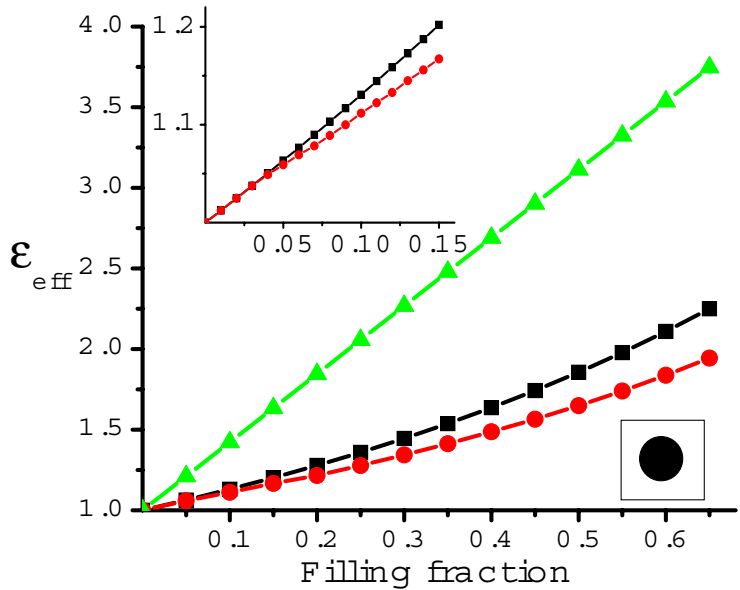}
\vspace{-0.9cm} \narrowtext \caption{In-plane effective dielectric
constant for the $H$-mode for uniaxial PC of solid graphite
cylinders with $\varepsilon_{\parallel} = 1.8225$ and
$\varepsilon_{\perp} = 5.226 $ in air, $\varepsilon_b = 1$
(circles). Straight line (triangles) is the effective dielectric
constant for the $E$-mode, Eq. (\ref{six}). The squares show the
results of the Maxwell-Garnett approximation (\ref{twentysix}).
Insert shows the region of small filling fractions.} \label{fig.1}
\end{center}
\end{figure}
To check the validity of the Maxwell-Garnett approximation, we plot
in Fig. \ref{fig.1} (squares) the effective dielectric constant
proposed in Ref. [\cite{Garcia}]
\begin{equation}
\varepsilon^{(H)}_{MG}=\frac{\varepsilon_{\parallel}+ \Delta +
f(\varepsilon_{\parallel}-\Delta)} {\varepsilon_{\parallel}+
\Delta - f(\varepsilon_{\parallel}-\Delta)}. \label{twentysix}
\end{equation}
Here $\Delta=\sqrt{\varepsilon_{\parallel}/\varepsilon_{\perp}}$.
One can see that for all filling fractions the Maxwell-Garnett
approximation gives overestimated values for the effective
dielectric constant. For a very dilute system, $f < 0.07$, the
Maxwell-Garnett approximation gives results that are practically
indistinguishable from the exact ones (See insert in Fig. \ref
{fig.1}.). For the close-packed array of cylinders the
Maxwell-Garnet approximation overestimates the dielectric constant
by about 25\%.

\begin{figure}
\begin{center}
\vspace{-0.2cm} \hspace {-1cm} \epsfxsize 8cm \epsfbox{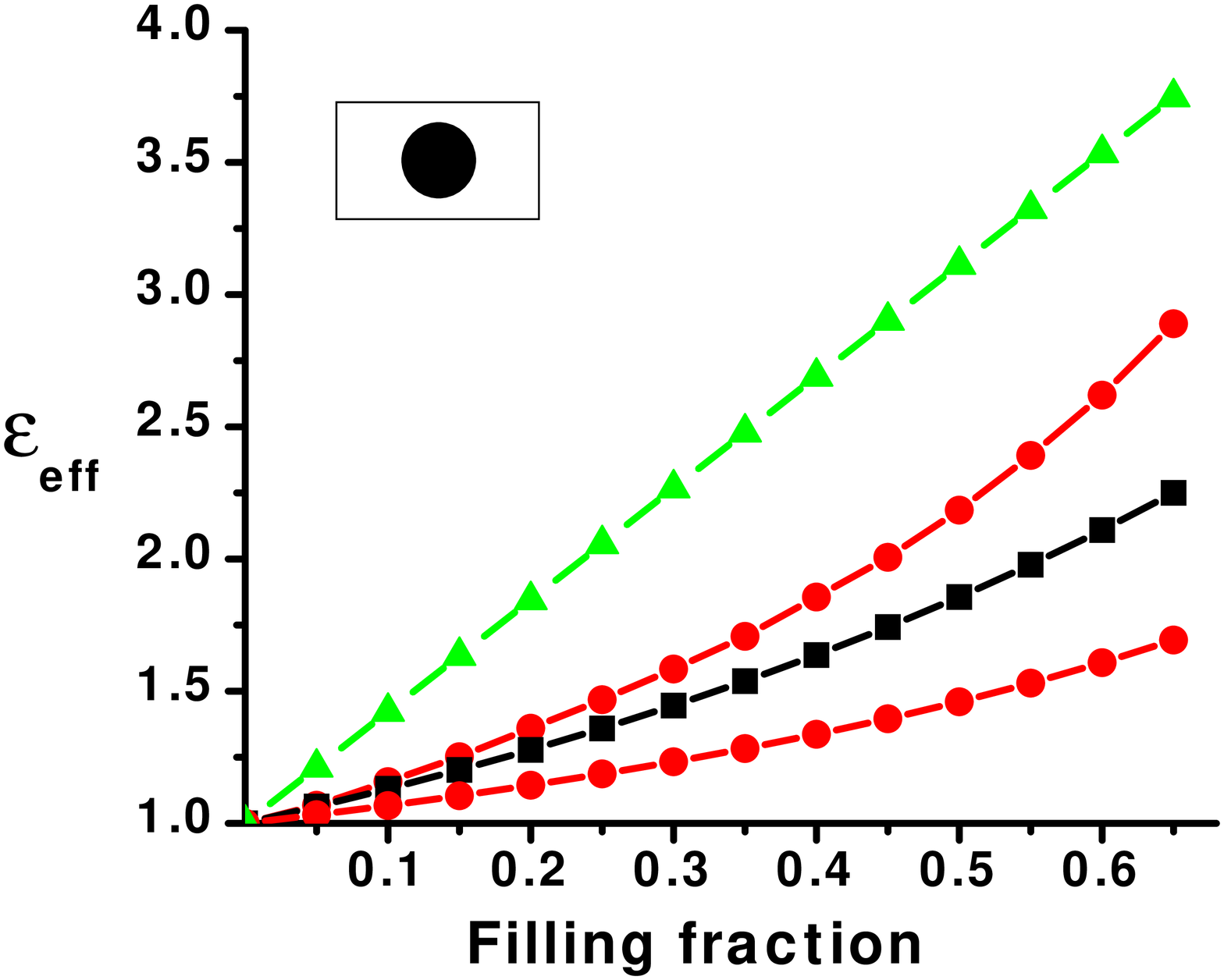}
\vspace{-0.2cm} \narrowtext \caption{ A plot of the principal
effective dielectric constants for the PC of solid carbon
cylinders arranged in a rectangular lattice. The larger (smaller)
dielectric constant $\varepsilon_{1}$ ($\varepsilon_{2}$)
corresponds to the direction of the vector $\bf E$ along the short
(long) side of the rectangle. The Maxwell-Garnett dielectric
constant is shown by the squares. } \label{fig.2}
\end{center}
\end{figure}

In Fig. \ref{fig.2} we plot two principal dielectric constants for
the biaxial PC of solid carbon cylinders with a rectangular unit
cell. The ratio of the sides of the rectangle is 1:2. The
difference between the two dielectric constants increases with the
filling fraction, giving rise to a higher anisotropy of the
corresponding effective medium. The Maxwell-Garnett approximation
Eq. (\ref{twentysix}), which does not take into account the
anisotropy of the unit cell, gives the values for
$\varepsilon_{MG}$ that lie between the two principal values,
$\varepsilon_{1} < \varepsilon_{MG} < \varepsilon_{2}$.

\section{Uniaxial photonic crystal of carbon nanotubes}

In our model we consider the carbon nanotubes as hollow graphite
cylinders. In the experimental study\cite{Heer} of the dielectric
properties of   carbon nanotues the outer radius of the cylinders
was approximately $R=5$ nm. The nanotubes formed a thin film and
they were oriented along a specific direction. Although the
nanotubes were not necessarily arranged periodically, one can
assume that they formed almost a regular lattice, since the
nanotube density is about 0.6 - 0.7 which is near the value of $f_c
= \pi/4 \approx 0.785$ for a close-packed structure.  Thus, the
separation between the nanotubes (the period of the square lattice
$d$) slightly exceeds $2R$, and in Ref. [\cite{Garcia}] it was
estimated to be $d=10.15$ nm. The inner radius $\gamma R = 0.25 -
2$ nm.\cite{Garcia} was evaluated from the amount of
electromagnetic absorption for the $E$-polarized light. The four
parameters $f$, $R$, $\gamma$, and $A_c = a^2$ are not independent
but related by the formula,
\begin{equation}
f =\pi R^2 (1-\gamma^2)/A_c.
\label{geom}
\end{equation}
Substituting the aforementioned parameters of the square unit cell
into this formula allows one to check that they are self-consistent. It is
worthwhile to mention that the background material in the
experiment\cite{Heer} is not air but the host material Delrin or
Teflon with $\varepsilon_b >1$. Since neither the density of the
host material nor its dielectric constant is known, one cannot
expect very good agreement between the experimental
results\cite{Heer} and theory. In all theoretical considerations
it was assumed that $\varepsilon_b =1$. Because of this lack of
experimental data, the effective medium theories\cite{Garcia,Lu}
and the results shown in Fig. \ref{fig.1} give lower values for
$\varepsilon_{eff}$ than that observed in the
experiment.\cite{Heer}
\begin{figure}
\begin{center}
\vspace{-0.2cm} \hspace {-0.5cm} \epsfysize 8cm \epsfbox{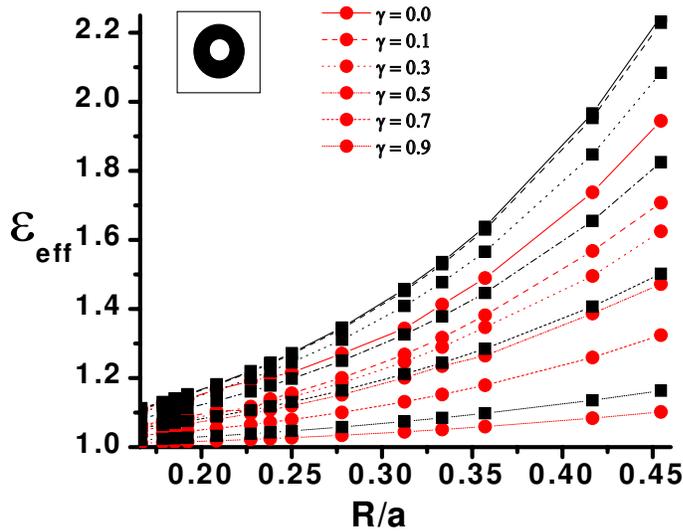}
\vspace{-0.9cm} \narrowtext \caption{The plot of the effective
dielectric constant for square lattice of carbon nanotubes versus
the outer radius for tubes with different ratios of the inner and
outer radii, $\gamma = 0.1, 0.3, 0.5, 0.7$. The exact results are
shown by circles and the results of the Maxwell-Garnett
approximation are shown by squares.} \label{fig.3}
\end{center}
\end{figure}

It is obvious that the inner cavity reduces the permittivity of an
isolated nanotube as compared to a solid graphite cylinder of the
same size. It was argued\cite{Garcia} that for a periodic arrangement the
effect of the inner cavity is less than that for a single cylinder
and even can be ignored, if the ratio between the inner and outer
radii $\gamma$ does not exceed 0.4. This conclusion
was supported by comparing the results of the Maxwell-Garnet
approximation Eq. (\ref{twentysix}) and numerical band structure
calculations. In Fig. {\ref{fig.3}} we plot the dielectric
constant for a square lattice of hollow carbon nanotubes and
compare the exact results obtained from Eqs. (\ref{twenty}),
(\ref{twentyone}), and (\ref{twentyfive}) (shown by the circles) with
the results given by the Maxwell-Garnett approximation (squares).
One can see that, for the same outer radius, the effective
dielectric constant drops with an increase of the inner radius.
Thus, if the outer radius is fixed, the dependence on the inner
radius cannot be ignored, even in the Maxwell-Garnett
approximation. However, the effective dielectric constants
exhibits much less sensitivity to the internal radius if it is
plotted as a function of the filling fraction, Fig. {\ref{fig.4}}.
\begin{figure}
\begin{center}
\vspace{-0.2cm} \hspace {-0.5cm} \epsfysize 8cm \epsfbox{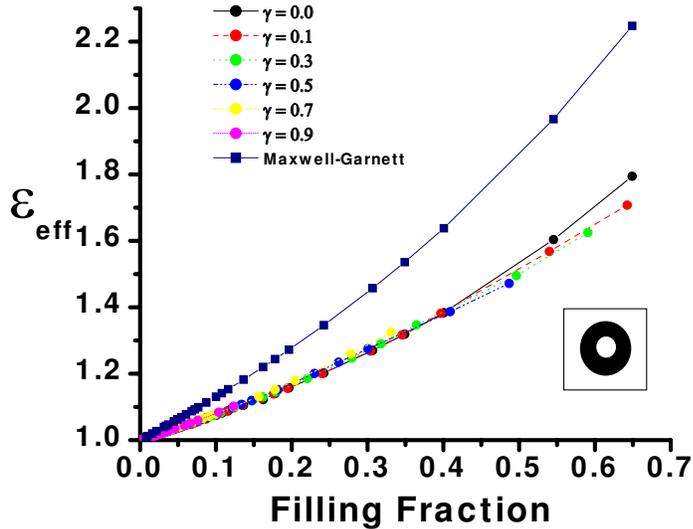}
\vspace{-0.9cm} \narrowtext \caption{A graph of the effective
dielectric constant for a square lattice of carbon nanotubes
versus filling fraction for tubes with different ratios of the
inner and outer radii, $\gamma = 0.1, 0.3, 0.5, 0.7$. The exact
results are shown by circles and the Maxwell-Garnett approximation
is shown by squares.} \label{fig.4}
\end{center}
\end{figure}
In the Maxwell-Garnet approximation (\ref{twentysix}) there is no
dependence on the parameter $\gamma$, therefore, this approximation
is represented by a single curve in Fig. {\ref{fig.4}}. Here, only
the total amount of the dielectric material is important, but not
the topology of the cylinders. In the exact theory the effective
dielectric constant depends on the details of the microstructure
of the photonic crystal, but as far as the filling fraction is
concerned, the topology plays a much less important role. Since the
cylinder is uniquely determined by either two parameters out of
three, $R$, $\gamma$, and $f$, the curves in Fig. {\ref{fig.4}}
may cross each other. This means that at the crossing point the
values of $f$ and $\gamma$ correspond to the same hollow cylinder.
This can be easily seen from Eq. (\ref{geom}).

\section{Conclusions }

We calculated the low-frequency dielectric tensor for 2D photonic
crystal of anisotropic parallel cylinders arranged in a periodic
lattice in the perpendicular plane. The exact
analytical formula for the principal values of the dielectric
tensor was obtained. The results are applied for the periodic arrangement of
carbon nanotubes which are rolled up from uniaxial graphite
crystal with static values of the dielectric tensor
$\varepsilon_{\parallel} = 1.8225$ and $\varepsilon_{\perp} =
5.226$. It was shown that the interior (vacuum) region of the nanotubes
has a small effect on the dielectric properties of the photonic
crystal and can be ignored. Although we are interested in the
static dielectric tensor, it is clear that the developed
long-wavelength limit approach remains valid, even for optical
frequencies since the period of the lattice of carbon nanotubes
$d=10$ nm is much less than the optical wavelength $\lambda
\approx 500$ nm. To calculate the dynamic dielectric tensor, one
has to substitute in the general formula Eq. (\ref{nineteen}) the
corresponding frequency-dependent values for
$\varepsilon_{\parallel}$ and $\varepsilon_{\perp} $. Of course at
finite frequencies Eq. (\ref{nineteen}) gives the real part of the
dielectric function. Calculations of the imaginary part require a
generalization of the presented theory. This result will be reported
elsewhere.

The exact theory presented here allows a calculation of the
effective dielectric constant of carbon nanotubes imbedded in a
gas. Due to high absorbability of nanotubes, the concentration of
gas in the interior region of the nanotubes may be different from that
in the atmosphere. This leads to slightly different dielectric
constants of the material in the interior and exterior regions of
the cylinders. This effect can be registered by precise
measurements of the shift of the resonant frequency of a resonant
cavity.\cite{conf} Thus, the proposed theory may find applications in the
microwave detection of poisson gases in the atmosphere.

One more interesting application of carbon nantonube photonic
crystal is related to its huge anisotropy of the effective
dielectric constant. Recently Artigas and Torner\cite{Tor}
demonstrated that the electromagnetic surface wave (Dyakonov
wave\cite{Dyak}) can propagate along the surface of a  photonic
crystal with high optical anisotropy. This wave propagates in a
lossless dielectric medium and decays much slower than the surface
plasmon. Since crystals with huge optical anisotropy are rare in
nature, carbon nanotube photonic crystals may be considered as a
promising material for integrated photonic circuits.

\section{Acknowledgement}
This work is supported by CONACyT (Mexico) Grant No. 42136-F.


\end{document}